\documentclass[showpacs,aps,graphicx,twocolumn]{revtex4}

\usepackage{graphicx}
\usepackage{dcolumn}
\usepackage{amsmath}
\usepackage{relsize}

\begin{document}



\title{ Quantum information processing on nitrogen-vacancy ensembles with the local resonance assisted by circuit QED}

\author{Ming-Jie Tao, Ming Hua,  Qing Ai \footnote{Corresponding author: aiqing@bnu.edu.cn}, and Fu-Guo Deng}

\address{Department of Physics, Applied Optics Beijing Area Major Laboratory,
Beijing Normal University, Beijing 100875, China}

\date{\today}

\begin{abstract}
With the local resonant interaction between a
nitrogen-vacancy-center ensemble (NVE) and a superconducting
coplanar resonator, and the single-qubit operation, we propose two
protocols for the state transfer between two remote NVEs and for fast
controlled-phase (c-phase) on these NVEs, respectively. This hybrid
quantum system is composed of two distant NVEs coupled to separated
high-$Q$ transmission line resonators (TLRs), which are
interconnected by a current-biased Josephson-junction
superconducting phase qubit. The fidelity of our state-transfer
protocol is about $99.65\%$ within the operation time of $70.60$ ns.
The fidelity of our c-phase gate is about $98.23\%$ within the
operation time of $93.87$ ns. Furthermore, using the c-phase gate,
we construct a two-dimensional cluster state on NVEs in $n\times n$
square grid based on the hybrid quantum system for the one-way
quantum computation. Our protocol may be more robust, compared with
the one based on the superconducting resonators, due to the long
coherence time of NVEs at room temperature.
\end{abstract}


\maketitle

\section{Introduction}
\label{sec:Introduction}

Universal quantum logic gates \cite{Sleator,Nielsen1} are the key
element for a quantum computer. In recent decades, much attention
has been focused on the construction of universal quantum logic
gates with different physical systems, such as ion trap
\cite{Cirac1,Poyatos}, cavity quantum electrodynamics (QED)
\cite{cavity1,cavity2,fidlifuli}, nuclear magnetic resonance
\cite{NMRJones,NMRLong}, quantum dots
\cite{Loss,electron1,dotweisr,dothybrid}, photons with one  degree
of freedom (DOF) \cite{photon1,photon2} or two DOFs (that is, the
hyper-parallel photonic quantum computation)
\cite{hypercnotlpl,hypercnotsr,hypercnotPRA}, superconducting qubit
\cite{Makhlin2,Yang4,SQ1,SQ2,AiSQ}, circuit QED
\cite{circuitQED1,circuitQED2,circuitQED3,circuitQED4,circuitQED5,circuitQED6,circuitQED7},
microwave-photon resonators \cite{mwphoton,Hua,HuaSR}, and diamond
nitrogen-vacancy (NV) centers \cite{NVQC1,NVQC2}. Among the above
schemes, much attention has been paid to the generation of the
controlled-phase (c-phase) gate which can be used to realize
universal quantum computation assisted with single-qubit operations.

In order to realize scalable quantum computation, tunable coupling
and coherence time are of special importance. In this regard, each
quantum system has its own advantages and disadvantages, e.g., easy
operability but not enough long coherence time and thus
insufficiently high fidelity. In order to overcome the disadvantages
of each system to realize universal quantum computation, a hybrid
quantum system \cite{Xiang}, which is composed of two or more kinds
of quantum systems, has attracted much attention recently.

The hybrid systems composed of superconducting circuits and the
other quantum systems \cite{Xiang}, such as atoms
\cite{rensen,Deng}, molecules \cite{Rabl,Tordrup1}, spins
\cite{Imamo,Chen1,Bushev}, and solid-state devices
\cite{Zhang1,You1}, have been studied. As a result of long coherence
time of the NV-center spin \cite{Jelezko}
and the strong coupling between nitrogen-vacancy-center ensemble
(NVE) and superconducting resonator \cite{Kubo1,Ams,Sandner}, the
hybrid system composed of diamond NVE and superconducting circuit
plays a good platform for quantum information processing. Recently,
a lot of theoretical and experimental works have been done in the
quantum information processing based on the hybrid system
\cite{Kubo1,Sandner,Kubo2,Yang2,Yang1}. For example, in 2010, Kubo
and coworkers \cite{Kubo1} realized the strong coupling of a spin
ensemble, which is composed of NV-centers in a diamond crystal, to a
superconducting resonator. In 2012, Sandner \emph{et al}.
\cite{Sandner} showed that  a dense NVE  can be coupled to a high-Q
superconducting resonator at low temperature both in experiment and
in theory. In 2011,  Kubo \emph{et al}. \cite{Kubo2} reported the
experimental realization of a hybrid quantum circuit combining a
superconducting transmon qubit and an NVE. Yang \emph{et al}.
\cite{Yang1} studied the high-fidelity quantum memory in a hybrid
quantum computing system composed of an NVE and a current-biased
Josephson-junction superconducting phase qubit in a transmission
line resonator (TLR) (as the quantum data bus). They also
\cite{Yang2} presented a potentially practical proposal for creating
entanglement of two distant NVEs coupled to separated TLRs
interconnected by a current-biased Josephson-junction
superconducting phase qubit. In 2012, Chen, Yang, and Feng
\cite{fengmang} proposed a scheme for the state transfer between
distant NVEs coupled with a superconducting flux qubit each, by modulating
the coupling strength between flux qubits and that between a flux
qubit and an NVE.

\begin{figure*}[tpb]
\begin{center}
\includegraphics[width=7.6 cm,angle=270]{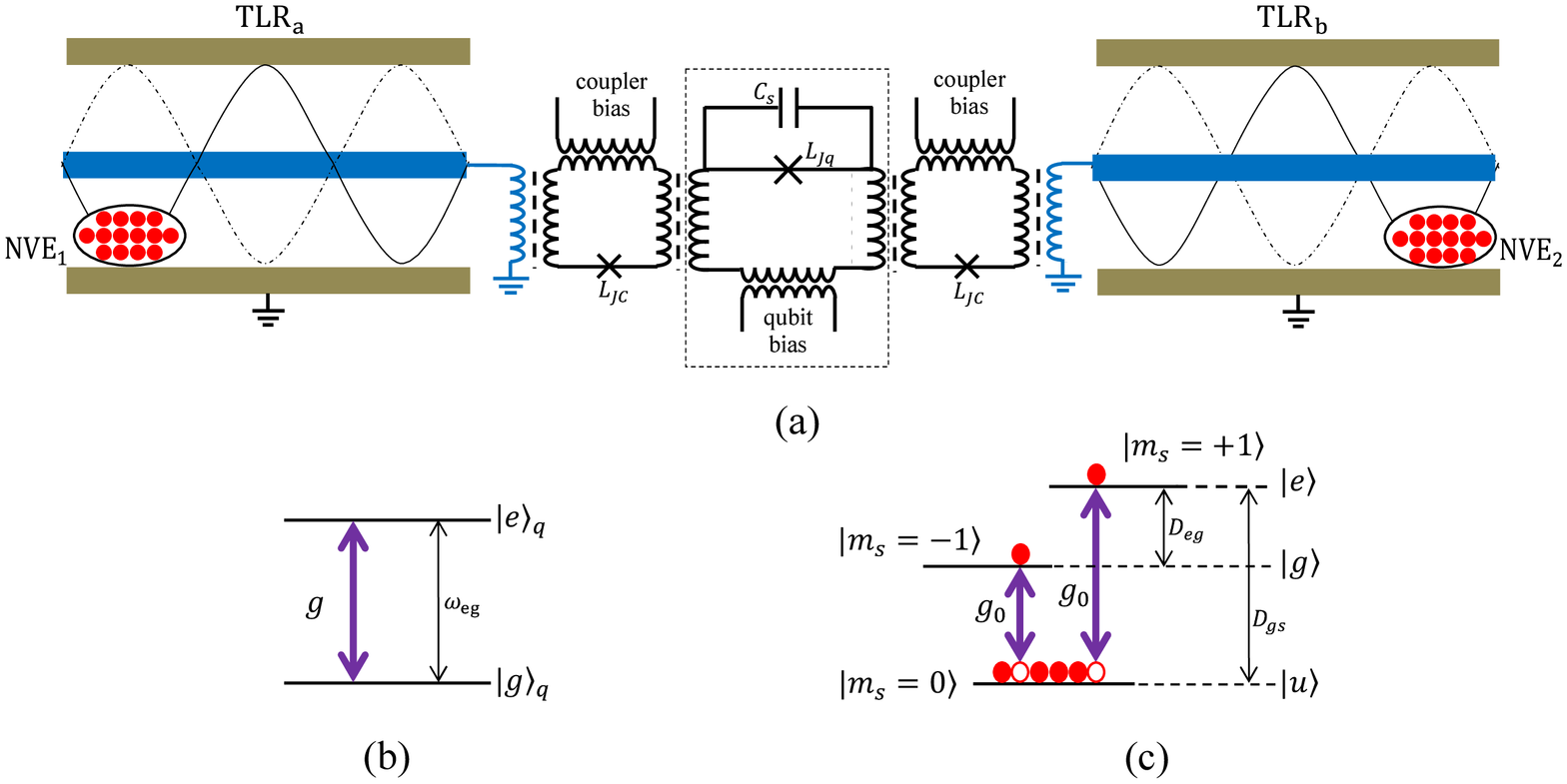}
\caption{(Color online) (a) Schematic diagram of the hybrid quantum
system for our quantum information processing on two NVEs. The TLRs
are connected to the SPQ by two couplers. The coupling strength
between the TLR and the SPQ can be tuned by the coupler. The NVEs
interact with the quantized fields of TLRs, which act as a quantum
data bus. (b) Level scheme of an SPQ. The SPQ is approximated as a
two-level system and there is a large energy gap $\omega_{eg}$ between
the two levels $|e\rangle_q$ and $\vert g\rangle_q$. (c) The
detailed energy configuration of a single NV-center under an
external magnetic field $\vec{B}$. The energy level difference
between $\vert m_{s}=\pm1\rangle$ is $D_{eg}=\gamma_{e}B$, where $\gamma_{e}$ is
the electron gyromagnetic ratio. And the transition frequency
between $\vert m_{s}=0\rangle$ and $\vert m_{s}=-1\rangle$ is
$D_{gs}-D_{eg}$.}\label{figure1}
\end{center}
\end{figure*}

In this paper, we consider quantum information processing in a
hybrid system composed of two distant NVEs coupled to separated
high-$Q$ TLRs, which are interconnected by a current-biased
Josephson-junction superconducting phase qubit (SPQ). By using the
resonant interaction between the resonator and the NVE with the
transition of $|m_s=0\rangle\leftrightarrow|m_s=-1\rangle$, and the
single-qubit operation, we propose a protocol for the quantum state
transfer between the two distant NVEs, and construct the c-phase and
CNOT gates on these NVEs as well. Because both the resonant
interaction between the NVE and the resonator and the single-qubit
rotation on NVEs are fast quantum manipulation, our state transfer
and gates have the features of a high fidelity and a short operation
time. The fidelity of our state tranfer and c-phase gate are about
$99.65\%$ and $98.23\%$, respectively. And the operation times of
them are $70.60$ ns and $93.87$ ns, respectively. Furthermore, we construct a
two-dimensional $n\times n$ squared grid based on the hybrid quantum
system interconnected by the SPQs. Thus, by virtue of the long
coherence time of NVE \cite{Yang2,Neumman}, we engineer a cluster
state of two-dimensional network for the one-way quantum computation
with the promising advantage compared with the one based on the
superconducting resonators \cite{Wu}.

\section{Model and Quantum Dynamics Of System}
\label{sec:Model}

Let us consider a hybrid quantum device composed of two distant NVEs
coupled to separated high-$Q$ TLRs as shown in
Fig.~\ref{figure1}(a). The two TLRs are interconnected by an SPQ.
The TLR with inductance $L$ and capacitance $C$ can be modeled as a
simple harmonic oscillator \cite{circuitQED1,circuitQED3} consisting
of a narrow center conductor and two nearby lateral ground planes
\cite{Yang1,Yang2}. The Hamiltonians of TLR$_a$ and TLR$_b$ can be
formed as
\begin{eqnarray}   
H_{a} &=& \omega_{a}a^{\dag}a
\end{eqnarray}
and
\begin{eqnarray}   
H_{b} &=& \omega_{b}b^{\dag}b,
\end{eqnarray}
respectively, where $a^{\dag}$ ($\omega_{a}=1/\sqrt{LC}$) and
$b^{\dag}$ ($\omega_{b}=1/\sqrt{LC}$) are the creation operators
(transition frequencies) of TLR$_a$ and TLR$_b$, respectively.

The circuit in the dashed-line box of Fig.~\ref{figure1}(a) is an SPQ. With the lowest two energy levels of an SPQ, the Hamiltonian is
\begin{equation}   
H_{q}=\frac{1}{2}\omega_{eg}\sigma_{z}.
\end{equation}
Here $\omega_{eg}$ is the resonant transition frequency between the
two levels of the SPQ (see Fig.~\ref{figure1}(b)), which can be
changed by the external flux bias to the qubit
\cite{Makhlin2,Galiautdinov}. $\sigma_{z}=|e\rangle\langle
e|-|g\rangle\langle g|$ is the Pauli spin operator of the SPQ, where
$|g\rangle$ and $|e\rangle$ are the ground and excited states,
respectively. By means of couplers, two TLRs are indirectly coupled
to the SPQ and the coupling strength can be changed by applying
different flux to the coupler \cite{Allman}.

Taking the rotating-wave approximation into account, the interaction
Hamiltonians between TLRs and SPQ are
\begin{eqnarray}    
H_{aq} &=& g_{a}(a\sigma^{+}+a^{\dag}\sigma^{-}),
\end{eqnarray}
and
\begin{eqnarray}    
H_{bq} &=& g_{b}(b\sigma^{+}+b^{\dag}\sigma^{-}),
\end{eqnarray}
respectively. Here, $g_{a}=(g_{b}=)g$ is the coupling strength
between TLR$_a$ (TLR$_b$) and SPQ. $\sigma^{+}=|e\rangle\langle g|$
$(\sigma^{-}=|g\rangle\langle e|)$ is the raising (lowering)
operator of the SPQ.

NV-centers in the device possess a V-type three-energy-level
configuration as shown in Fig.~\ref{figure1}(c). Every NV-center is
negatively charged with two unpaired electrons located at the
vacancy. Thus, the spin-spin interaction leads to the same energy
splitting between $|m_{s}=0\rangle$ and $|m_{s}=\pm1\rangle$, i.e.,
$D_{gs}=2.88$  GHz \cite{Hanson}. When there is an external magnetic
field $\vec{B}$ along the NV-center symmetry axis, the degeneracy of
the levels $|m_{s}=\pm1\rangle$ is lifted, which causes a level
splitting $D_{eg}=\gamma_{e}B$,  with $\gamma_{e}$ being the
gyromagnetic ratio of electron \cite{Chen1}. For simplicity, we
label the states of the NV-center $|m_{s}=0\rangle,
|m_{s}=-1\rangle$, and $|m_{s}=1\rangle$ as $|u\rangle, |g\rangle$,
and $|e\rangle$, respectively. Moreover, the lowest level
$|u\rangle$ of the NV-center is an auxiliary state in the present
work. There are $N$ NV-centers in the single NVE and the Hamiltonian
of an NVE reads
\begin{equation}    
H_{k}=\frac{1}{2}\omega_{k,0}S_{k,0}^{z}+\frac{1}{2}\omega_{k,1}S_{k,1}^{z},
\end{equation}
where $k=1,2$ are on behalf of NVE$_1$ and NVE$_2$.
$\omega_{k,0}=D_{gs}-\gamma_{e}B_{k}$ and
$\omega_{1,1}=\omega_{2,1}=D_{gs}$ are the transition frequencies of
$|u\rangle\leftrightarrow|g\rangle$ and
$|u\rangle\leftrightarrow|e\rangle$, respectively.
$S_{k,0}^{z}=\sum_{i=1}^N \tau_{k,i}^z$ and
$S_{k,0}^{\pm}=\sum_{i=1}^N \tau_{k,i}^{\pm}/\sqrt{N}$ are a set of
collective spin operators \cite{Song,Ai2} for NVE $k$ with
$\tau_{k,i}^z=|g\rangle_{k,i}\langle g|-|u\rangle_{k,i}\langle u|$,
$\tau_{k,i}^{+}=|g\rangle_{k,i}\langle u|$, and
$\tau_{k,i}^{-}=|u\rangle_{k,i}\langle g|$. And
$S_{k,1}^{z}=\sum_{i=1}^N v_{k,i}^z$ and $S_{k,1}^{\pm}=\sum_{i=1}^N
v_{k,i}^{\pm}/\sqrt{N}$ are the other set of collective spin
operators for  NVE $k$ with $\upsilon_{k,i}^z=|e\rangle_{k,i}\langle
e|-|u\rangle_{k,i}\langle u|$,
$\upsilon_{k,i}^{+}=|e\rangle_{k,i}\langle u|$, and
$\upsilon_i^{-}=|u\rangle_{k,i}\langle e|$.

The NVE qubit in this work is encoded in the following $|0\rangle$
and $|1\rangle$ states
\begin{eqnarray}    
|0\rangle_{k} &=& S_{k,0}^{+}|U\rangle_k=\frac{1}{\sqrt{N}}\sum_{i=1}^{N}|u_{1}\cdots g_{i}\cdots u_{N}\rangle_{k},  \\
|1\rangle_{k} &=& S_{k,1}^{+}|U\rangle_k=\frac{1}{\sqrt{N}}\sum_{i=1}^{N}|u_{1}\cdots e_{i}\cdots u_{N}\rangle_{k},
\end{eqnarray}
where $|U\rangle_{k}=|u_{1}\cdots u_{i}\cdots u_{N}\rangle_{k}$ is
the auxiliary state for an NVE. Using the rotating-wave
approximation, the interaction Hamiltonian of an NVE coupled to the
corresponding TLR by the magnetic-dipole coupling reads \cite{Yang2}
\begin{eqnarray}    
H_{a1} &=& g_{1}(S_{1,0}^{+}a+S_{1,0}^{-}a^{\dag}+S_{1,1}^{+}a+S_{1,1}^{-}a^{\dag}),
\end{eqnarray}
and
\begin{eqnarray}    
H_{b2} &=& g_{2}(S_{2,0}^{+}b+S_{2,0}^{-}b^{\dag}+S_{2,1}^{+}b+S_{2,1}^{-}b^{\dag}),
\end{eqnarray}
where $g_{1}=\sqrt{N}g_{0}$, $g_{2}=\sqrt{N}g_{0}$, and $g_{0}$
is the single NV-center vacuum Rabi frequency. When the NVE is
placed near the field antinode, the spatial dimension of the
ensemble is smaller than the mode wavelength so that the spins in
the NVE interact quasi-homogeneously with a single mode
electromagnetic field.

The total Hamiltonian of our hybrid device composed of two NVEs
coupled to separated TLRs interconnected by an SPQ can be described
as
\begin{equation}    
H=H_{a}+H_{b}+H_{q}+H_{1}+H_{2}+H_{aq}+H_{bq}+H_{a1}+H_{b2}. \label{eq:OriginalH}
\end{equation}
In the interaction picture, by assuming $\omega_{a}=\omega_{eg}=\omega_{b}$, the total Hamiltonian becomes
\begin{eqnarray}\label{state12}    
H_{I} \!\!\! &=& \!\!\! g(a^{\dag}\sigma^{-}+b^{\dag}\sigma^{-})+g_{1}(a^{\dag}S_{1,0}^{-}e^{-i\delta_{1,0}t}+a^{\dag}S_{1,1}^{-}e^{-i\Delta t})   \nonumber \\
     &&  +g_{2}(b^{\dag}S_{2,0}^{-}e^{-i\delta_{2,0}t}+b^{\dag}S_{2,1}^{-}e^{-i\Delta t})+\textrm{h.c.}
\end{eqnarray}
Here, $\delta_{1,0}=\omega_{1,0}-\omega_{a}$, $\delta_{2,0}=\omega_{2,0}-\omega_{b}$, and $\Delta=\omega_{1,1}-\omega_{a}=\omega_{2,1}-\omega_{b}$.

\begin{table}
\caption{Scheme for the quantum state transfer between two NVEs.}
\begin{tabular}{l c c c}
\hline \hline ~~~~Step          & Transition & Coupling &~ Pulse \\
\hline
1)~Rotate NVE$_1$ & $|1\rangle_{1} \rightarrow |U\rangle_{1}$                                                      &  $\Omega_R/2$ &  $\pi$  \\
2)~Resonate       & $|0\rangle_{1}|0\rangle_{a} \rightarrow |U\rangle_{1}|1\rangle_{a}$                            &  $g_{1}$      &  $\pi$  \\
3)~Resonate       & $|1\rangle_{a}|g\rangle_{q}|0\rangle_{b} \rightarrow |0\rangle_{a}|g\rangle_{q}|1\rangle_{b}$  &  $g$          &  $\sqrt{2}\pi$  \\
4)~Resonate       & $|1\rangle_{b}|U\rangle_{2} \leftrightarrow |0\rangle_{b}|0\rangle_{2}$                        &  $g_{2}$      &  $\pi$  \\
5)~Rotate NVE$_2$ & $|U\rangle_{2} \rightarrow |1\rangle_{2}$                                                      &  $\Omega_R/2$ &  $3\pi$ \\
\hline
\hline
\end{tabular}\label{Table0}
\end{table}

\section{Quantum State Transfer between NVEs}
\label{sec:QST}

In quantum information processing, the transfer of the quantum state
from one location to another is an important task and it is the
premise of the realization of large-scale quantum computing and
quantum networks.  Our device for the state transfer between distant
NVEs is shown in Fig.~\ref{figure1}(a) and this task can be achieved
with five steps shown in  Table~\ref{Table0}. Its principle can be
described in detail as follows.

Suppose that the hybrid quantum system composed of two NVEs, two
TLRs, and the SPQ for the state transfer is initially in the
superposition state
\begin{eqnarray}
|\phi\rangle_{I}=(\alpha|0\rangle_{1}+\beta|1\rangle_{1})
|0\rangle_{a}|g\rangle_{q}|0\rangle_{b}|U\rangle_{2},\label{phii}
\end{eqnarray}
where $\alpha$ and $\beta$ are complex numbers. $|n\rangle_{a}$ and
$|n\rangle_{b}$ indicate the Fock states of TLR$_a$ and TLR$_b$,
respectively.

\begin{figure*}[tpb] 
\begin{center}
\includegraphics[width=5.5cm,angle=0]{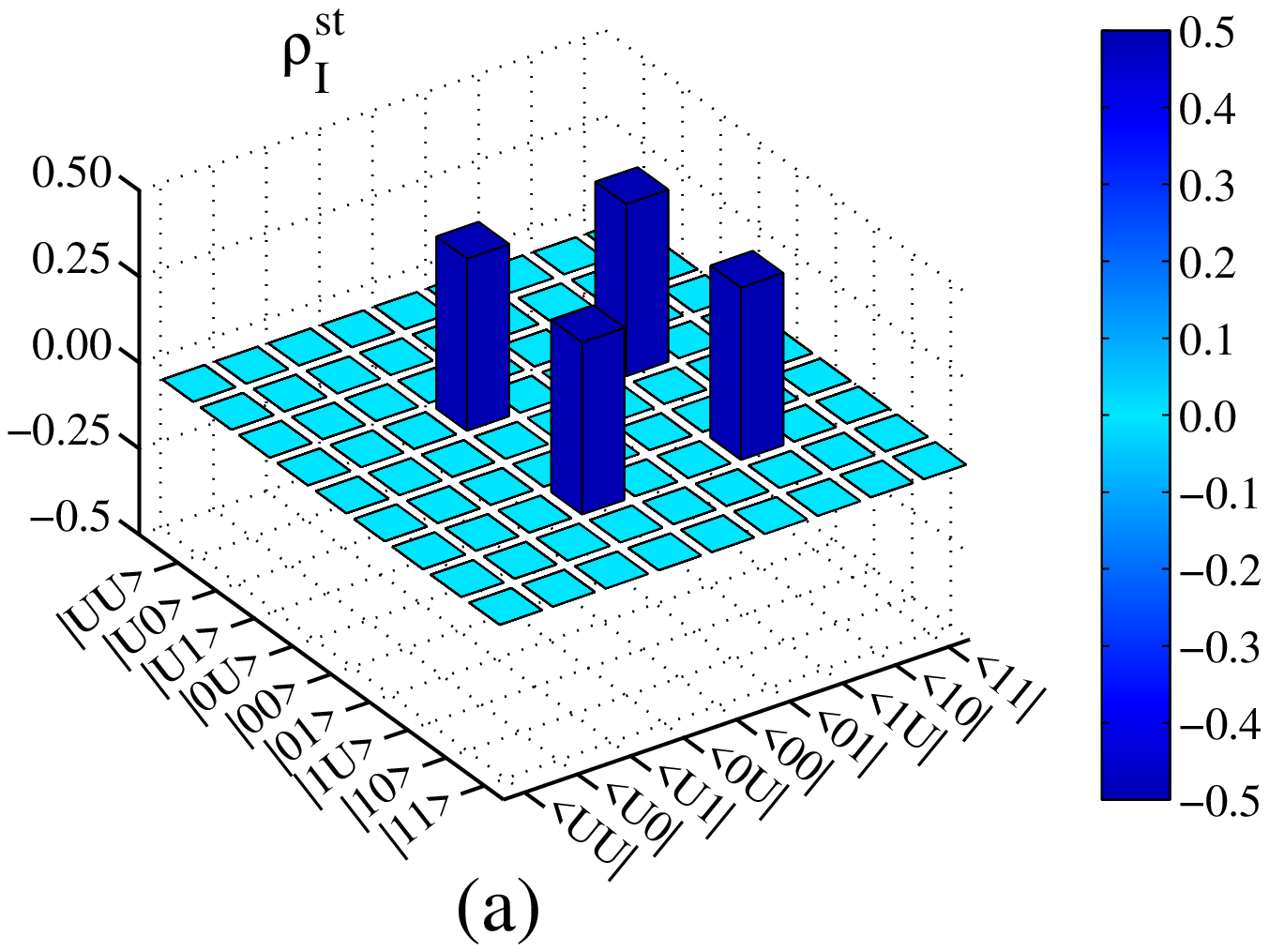}$\;\;\;\;\;\;\;\;\;\;\;\;$
\includegraphics[width=5cm,angle=0]{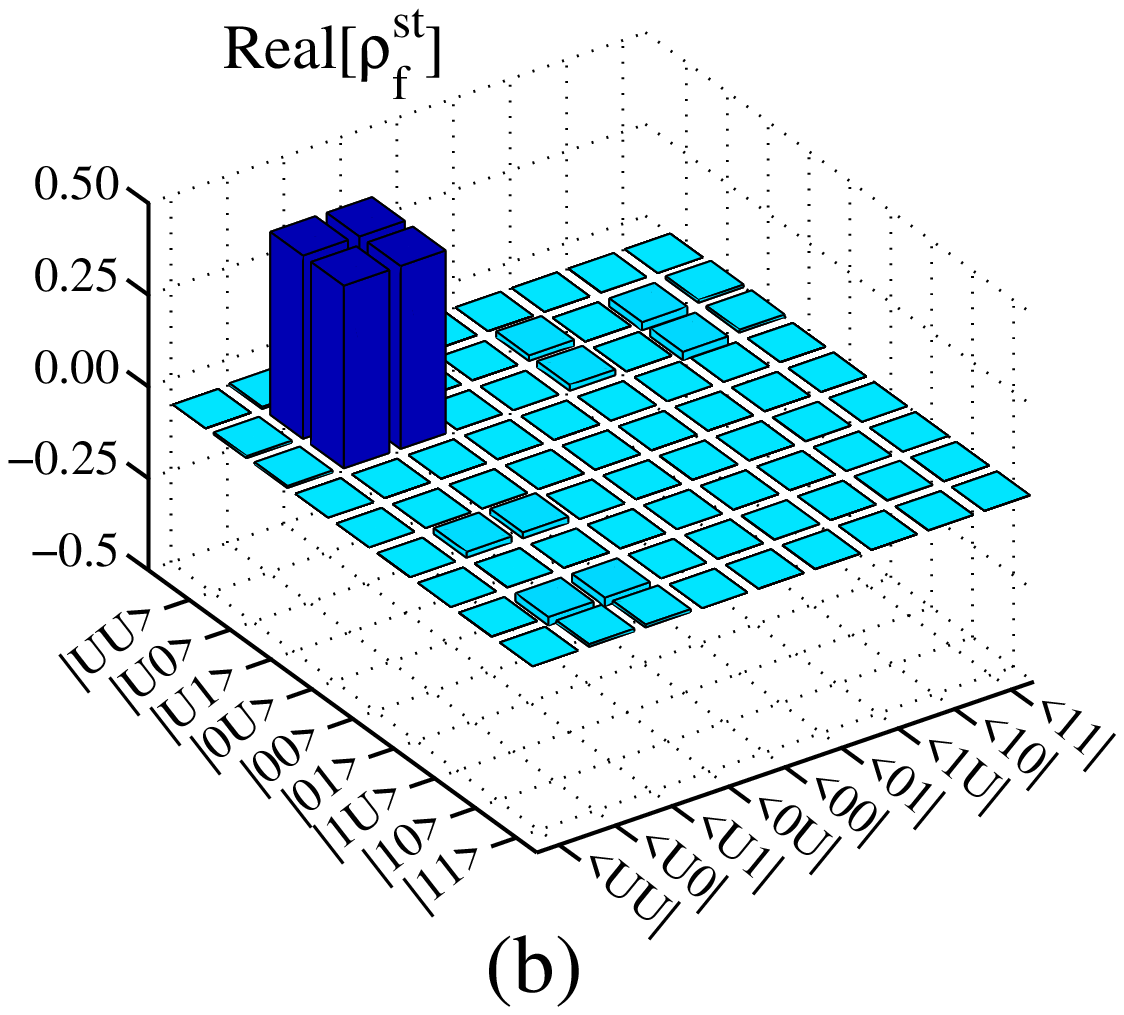}$\;\;\;\;\;\;\;$
\includegraphics[width=5cm,angle=0]{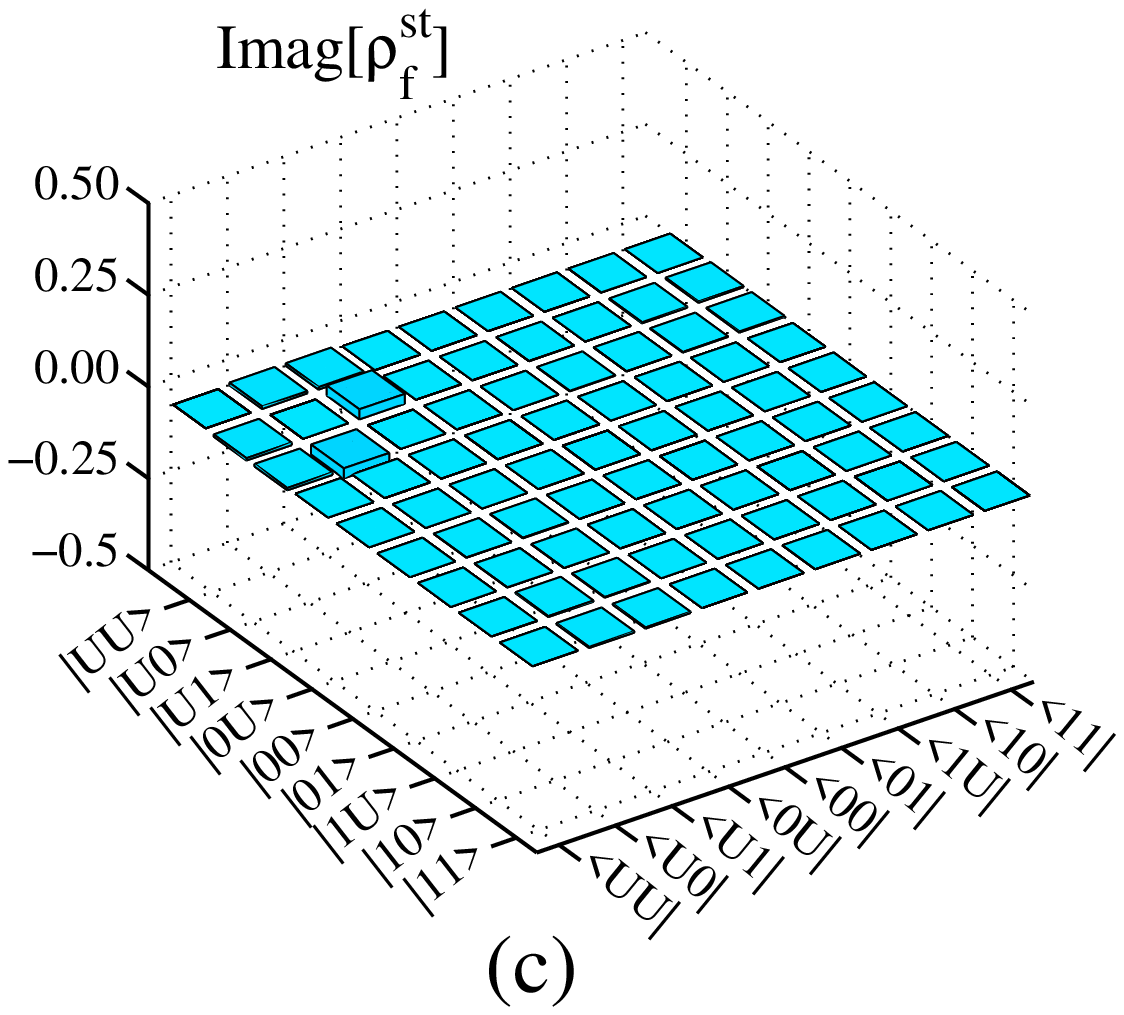}
\end{center}
\caption{(Color online) (a) The density matrix of the initial state
$\rho_I^{\text{st}}=\textrm{Tr}_{a,q,b}(\vert\phi\rangle_{I}\langle\phi\vert)$ (with
$\alpha=\beta=1/\sqrt{2}$) of the system. (b) the real and (c)
imaginary parts of the density matrix $\rho_f^{\text{st}}$ after the implementation of
the state transfer, respectively.}\label{figure2}
\end{figure*}

In step~(1), we  apply an external drive field governed by
$H_{D}^{\phi}=\Omega_{R}\exp(-i\omega_{d}t)S_{1,1}^{+}+\textrm{h.c.}$
with the Rabi frequency $\Omega_{R}$ and $\omega_{d}=\omega_{1,1}$
to flip the two states $|1\rangle_{1}\leftrightarrow |U\rangle_{1}$
of NVE$_1$. With the drive field, the Hamiltonian of the subsystem
composed of NVE$_1$ and TLR$_a$ is
\begin{equation}    
 H_{s1}^{\prime\phi}=H_{a}+H_{1}+H_{a1}+H_{D}^{\phi}.
\end{equation}
In the interaction picture, the Hamiltonian reads
\begin{equation}    
H_{s1}^{\phi}=g_{1}(a^{\dag}S^{-}_{1,0}e^{-i\delta_{1,0}t}+a^{\dag}S^{-}_{1,1}e^{-i\Delta t})+\frac{\Omega_{R}}{2}S_{1,1}^{+}+\textrm{h.c.}
\end{equation}
In the large-detuning regime $\delta_{1,0},\Delta\gg g_{1}$, under the rotating-wave approximation, the Hamiltonian reads
\begin{equation}    
H_{s1}^{\phi}\approx\frac{\Omega_{R}}{2}(S_{1,1}^{+}+S_{1,1}^{-}).
\end{equation}
When the drive field is applied on the NVE$_1$ for a duration $t=\pi/\Omega_{R}$, the evolution of NVE$_1$ follows
\begin{eqnarray}
|1\rangle_{1}\rightarrow -i|U\rangle_{1},
\end{eqnarray}
while the states of TLR$_b$ and SPQ remain unaltered. That is, the evolution of the state of the system is
\begin{eqnarray}    %
|\phi\rangle_{I}\;\rightarrow\;|\phi\rangle_{1} \!\!\! &=& \!\!\! (\alpha |0\rangle_{1}
-i\beta |U\rangle_{1})|0\rangle_{a}|g\rangle_{q}|0\rangle_{b}|U\rangle_{2}.\;\;\;\;\;\;
\end{eqnarray}

In step~(2), we tune the transition $|0\rangle_{1}\leftrightarrow
|U\rangle_{1}$ of NVE$_1$ to achieve its local resonance with
TLR$_a$ by adjusting the applied magnetic field $\vec{B_1}$, and
turn down the interaction between the SPQ and two TLRs by decreasing
the coupling strength $g$ to $0.5$ MHz $\ll \min(g_{1}=16$ MHZ,
$g_{2}=20$ MHz) \cite{Allman}. Due to the weak coupling strength
between the SPQ and TLRs, the energy transfer between the SPQ and
TLRs can be omitted. The interaction Hamiltonian of the subsystem
composed of NVE$_1$ and TLR$_a$ is given by
\begin{equation}
H_{2}^{\phi}=g_{1}(a^{\dag}S_{1,0}^{-}+a^{\dag}S_{1,1}^{-}e^{-i\Delta t}+\textrm{h.c.}).
\end{equation}
In the large-detuning regime $\Delta\gg g_{1}$, we can ignore the fast-oscillating terms, and the subsystem Hamiltonian can be simplified as
\begin{equation}
H_{2}^{\phi}\approx g_{1}(a^{\dag}S_{1,0}^{-}+aS_{1,0}^{+}).
\end{equation}
The evolution operator of this resonant interaction is written as $U_{2}(t)=\exp(-iH_{2}t)$. After a duration $t=\pi/2g_{1}$, we can obtain
\begin{eqnarray}
|0\rangle_{1}|0\rangle_{a}\rightarrow -i|U\rangle_{1}|1\rangle_{a}.
\end{eqnarray}
After this local resonance, the state of the total system becomes
\begin{eqnarray}
|\phi\rangle_{2} \!\!\! &=& \!\!\! |U\rangle_{1}(-i\alpha |1\rangle_{a}-i\beta |0\rangle_{a})|g\rangle_{q}|0\rangle_{b}|U\rangle_{2}.
\end{eqnarray}

In step~(3), we turn up the coupling between the SPQ and TLRs by increasing the coupling strength $g$ to $104$ MHz \cite{Allman}.
And we can tune the transition frequencies of NVEs to be largely detuned with TLRs. In this case,
there is only the energy transfer between the TLRs and the SPQ. The corresponding effective Hamiltonian is 
\begin{equation}    
H_{3}^{\phi}=g(a^{\dag}\sigma^{-}+b^{\dag}\sigma^{-}+\textrm{h.c.}).
\end{equation}
Governed by this Hamiltonian with the duration $t=\pi/\sqrt{2}g$, the system evolves from the state $|\phi\rangle_{2}$ to
\begin{eqnarray}
|\phi\rangle_{3} \!\!\! &=& \!\!\! |U\rangle_{1}|0\rangle_{a}|g\rangle_{q}(i\alpha |1\rangle_{b}-i\beta |0\rangle_{b})|U\rangle_{2}.
\end{eqnarray}

In step~(4), we tune the transition frequency $\omega_{2,0}$ between
$|0\rangle_{2}$ and $|U\rangle_{2}$ of NVE$_2$ to be equal to the
frequency $\omega_{b}$ of TLR$_b$ by adjusting the external magnetic
field $\vec{B_2}$, and turn down the interaction between TLRs and
the SPQ by turning the coupling strength $g$ to be $0.5$ MHz
\cite{Allman}. Without considering the weak interaction terms,  the
effective Hamiltonian is given by
\begin{equation}\label{state21}
\begin{split}
H_{4}^{\phi}&=g_{2}(b^{\dag}S_{2,0}^{-}+\textrm{h.c.}).
\end{split}
\end{equation}
In this step, the state driven by this Hamiltonian with the interval $t=\pi/g_{2}$ becomes
\begin{eqnarray}    
|\phi\rangle_{4} \!\!\! &=& \!\!\! |U\rangle_{1}|0\rangle_{a}|g\rangle_{q}|0\rangle_{b}(\alpha |0\rangle_{2}-i\beta |U\rangle_{2}).
\end{eqnarray}

In the last step~(5), we  apply a drive pulse with the duration
$t=3\pi/\Omega_{R}$   on NVE$_2$ to induce the transition between
$|1\rangle_{2}$ and $|U\rangle_{2}$. Thus an overall quantum state
transfer between NVE$_1$ and NVE$_2$ is implemented, leaving the
TLR$_a$, TLR$_b$, and the SPQ unchanged in the vacuum and ground
state, that is,
\begin{eqnarray}    
|\phi\rangle_{\textrm{F}} \!\!\! &=& \!\!\! |U\rangle_{1}|0\rangle_{a}|g\rangle_{q}|0\rangle_{b}(\alpha |0\rangle_{2}+\beta |1\rangle_{2}).
\end{eqnarray}

To show the feasibility of our proposal for the state transfer
between NVE$_1$ and NVE$_2$, we simulate the dynamics of the system
with the Hamiltonian shown in Eq.~(\ref{state12}). In the
simulations, we choose the parameters
$\omega_{a}/2\pi=\omega_{eg}/2\pi=\omega_{b}/2\pi=1.3$ GHz,
$\omega_{1,0}/2\pi=\omega_{2,0}/2\pi=1.73$ GHz for the
large-detuning case, $\omega_{1,1}/2\pi=\omega_{2,1}/2\pi=2.88$ GHz,
$g_{1}/2\pi=16$ MHz, $g_{2}/2\pi=20$ MHz, and $g/2\pi=104$ (0.5) MHz
when we turn up (down) the couplings between the SPQ and TLRs. The
Rabi frequency induced by the drive field is $\Omega_{R}/2\pi=50$
MHz.  If $\alpha=\sin\theta$ and $\beta=\cos\theta$, the final
(target) state is
$|\phi\rangle_{F}=|U\rangle_{1}|0\rangle_{a}|g\rangle_{q}|0\rangle_{b}(\sin\theta
|0\rangle_{2}+\cos\theta |1\rangle_{2})$. Here the average fidelity
of our proposal for the quantum state transfer is defined as
\cite{fidlifuli,HuaSR}
\begin{eqnarray}              
F_{\phi}=\frac{1}{2\pi}\int_{0}^{2\pi}\, _{F}\langle\phi|\rho_{f}^{\textrm{st}}|\phi\rangle_{F}
d\theta, \label{fidelity}
\end{eqnarray}
where  $\rho_{f}^{\textrm{st}}$ is the realistic density operator after our
state-transfer operation on the initial state $\left\vert
\phi\right\rangle_{I}$. Our simulation shows that the fidelity of
our state-transfer protocol is $99.65\%$ within the operation time
$70.60$ ns. Taking $\alpha=\beta=1/\sqrt{2}$ as an example, the
density operators of the initial state and the final state are shown
in Fig.~\ref{figure2}. The density
matrix is spanned in the basis
$\{~|U\rangle_{1}|U\rangle_{2},~|U\rangle_{1}|0\rangle_{2},~|U\rangle_{1}|1\rangle_{2},~|0\rangle_{1}|U\rangle_{2},~|0\rangle_{1}|0\rangle_{2}$,
$~|0\rangle_{1}|1\rangle_{2},~|1\rangle_{1}|U\rangle_{2},~|1\rangle_{1}|0\rangle_{2},~|1\rangle_{1}|1\rangle_{2}$\}.

\begin{figure*}[tpb] 
\begin{center}
\includegraphics[width=5.2cm,angle=0]{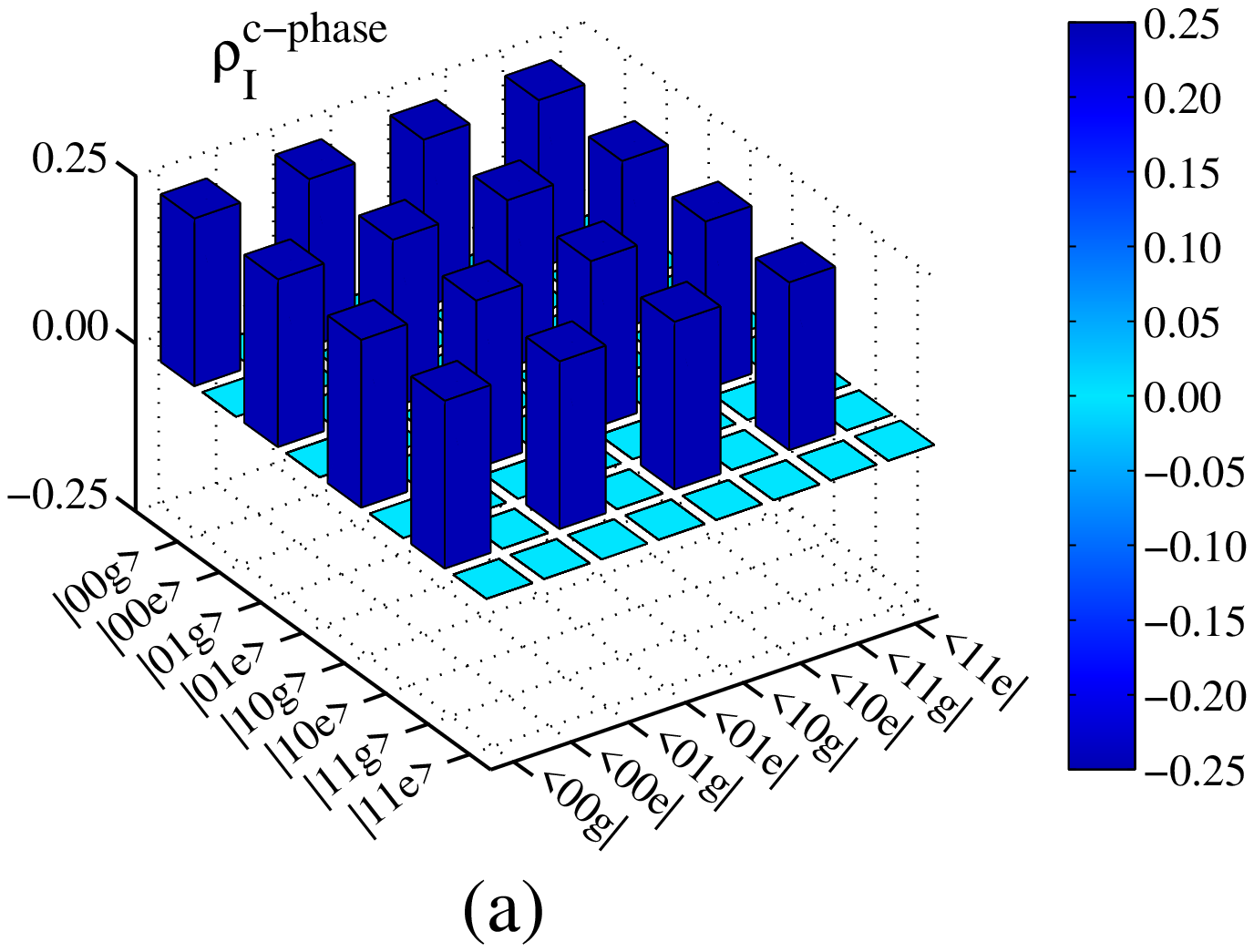}$\;\;\;\;\;\;\;\;\;$
\includegraphics[width=5.2cm,angle=0]{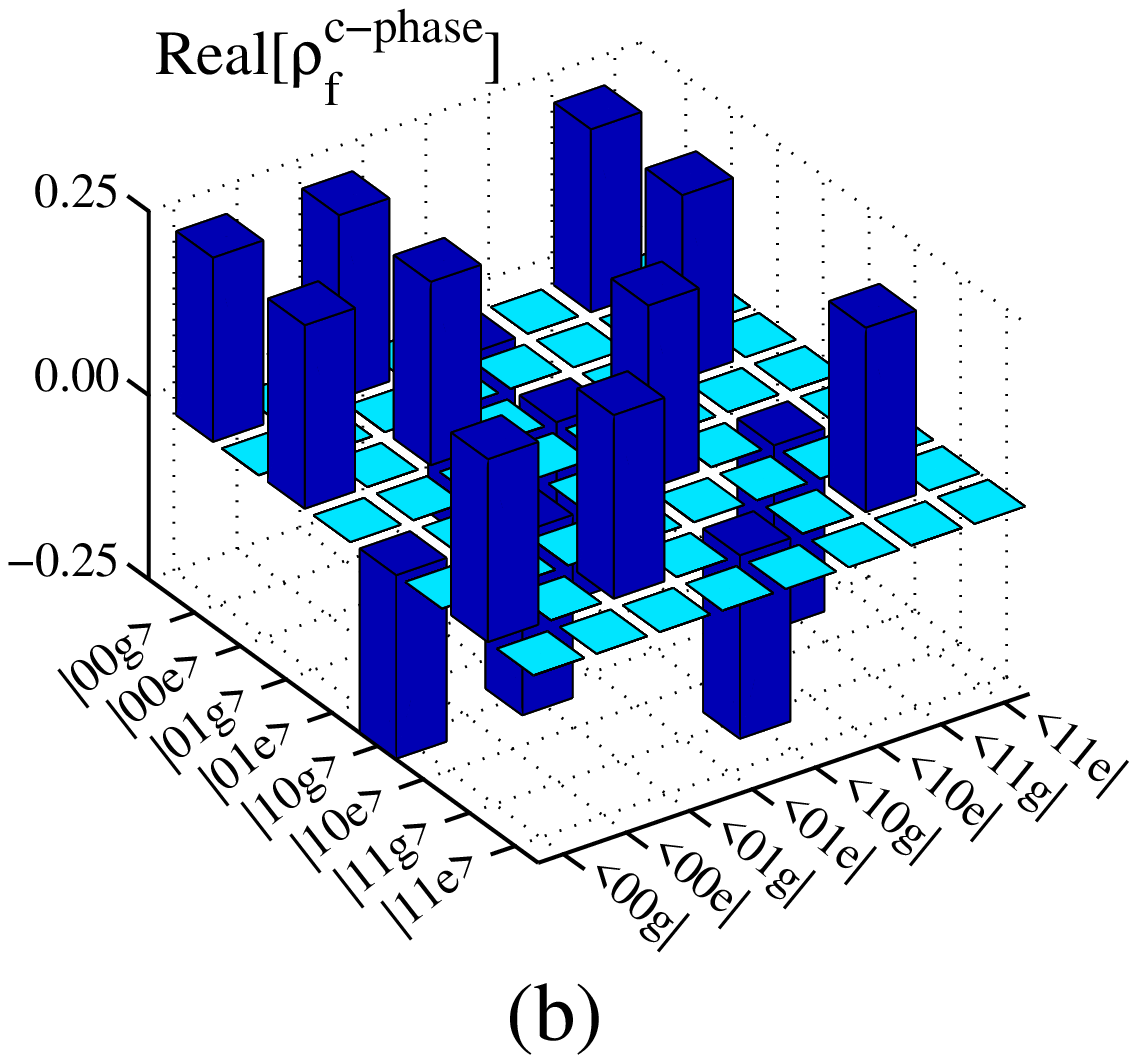}$\;\;\;\;\;$
\includegraphics[width=5.2cm,angle=0]{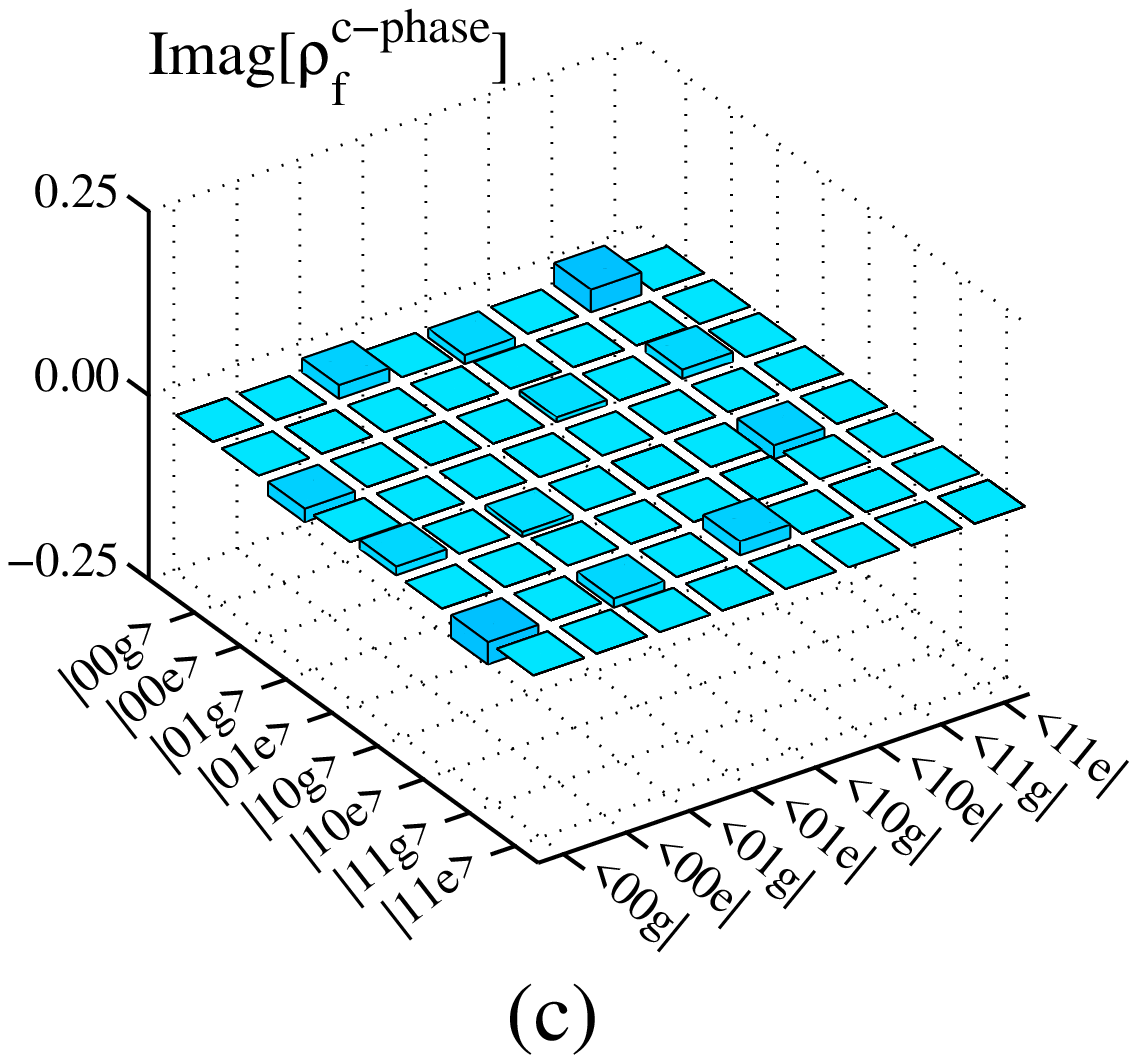}
\end{center}
\caption{(Color online) (a) The density matrix of the initial state $\rho_{I}^{\text{c-phase}}=\textrm{Tr}_{a,q,b}(\vert\psi\rangle_{I}\langle\psi\vert)$
(with $\alpha=\beta=\gamma=\delta=1/2$) of the system.
(b) the real and (c) imaginary parts of the density matrix $\rho_{f}^{\text{c-phase}}$ after the implementation of the c-phase gate, respectively.}\label{figure3}
\end{figure*}

\section{C-phase and CNOT Gates on two NVEs}
\label{sec:Gates}

C-phase gate is one of the significant quantum logic gates for
quantum information processing and it can be used to form a series
of universal gates to achieve quantum computation \cite{Nielsen1}
assisted by single-qubit operations. In the basis of two NVEs
$\{|0\rangle_{1}|0\rangle_{2},|0\rangle_{1}|1\rangle_{1},|1\rangle_{1}|0\rangle_{2},|1\rangle_{1}|1\rangle_{2}\}$,
the matrix of the c-phase gate reads
\[U_\textrm{c-phase} = \begin{pmatrix}
1 & 0 & 0 & 0 \\
0 & 1 & 0 & 0 \\
0 & 0 & -1 & 0 \\
0 & 0 & 0 & 1
\end{pmatrix}, \]
where there is a $\pi$ phase shift when the two-NVE system is in the
state $|0\rangle_{1}|1\rangle_{2}$. The initial state of the hybrid
quantum system composed of two NVEs, two TLRs, and an SPQ (the
device is shown in Fig.~\ref{figure1}(a)) is prepared as
\begin{eqnarray}    
|\psi\rangle_{I}
\!\!&=&\!\!(\cos\theta_1|0\rangle_{1}+\sin\theta_1|1\rangle_{1})(\cos\theta_2|0\rangle_{2}+\sin\theta_2|1\rangle_{2})
 \nonumber\\
                 && \otimes|g\rangle_{q}|0\rangle_{a}|0\rangle_{b}\nonumber\\
\!\!&=&\!\!(\alpha|0\rangle_{1}|0\rangle_{2}+\beta|0\rangle_{1}|1\rangle_{2}+\gamma|1\rangle_{1}|0\rangle_{2}+\delta|1\rangle_{1}|1\rangle_{2})\;\;\;\; \nonumber\\
                 && \otimes|g\rangle_{q}|0\rangle_{a}|0\rangle_{b},
\end{eqnarray}
where $\alpha=\cos\theta_1\cos\theta_2$,
$\beta=\cos\theta_1\sin\theta_2$, $\gamma=\sin\theta_1\cos\theta_2$,
and $\delta=\sin\theta_1\sin\theta_2$ are complex numbers. By
combining the single-qubit flip on NVEs and the resonant
interactions between NVEs and TLRs and those between TLRs and the
SPQ, the c-phase gate on NVE$_1$ and NVE$_2$ can be achieved by five
steps displayed in Table~\ref{Table1}.

\begin{table}
\caption{Scheme for the c-phase gate between two NVEs.}
\begin{tabular}{l c c c}
\hline \hline ~~~~~~Step          & Transition & Coupling & ~Pulse
\\ \hline
1)~Resonate         & $|0\rangle_{1}|0\rangle_{a} \rightarrow |U\rangle_{1}|1\rangle_{a}$                            &  $g_{1}$      &  $\pi$  \\
~~~\,Rotate NVE$_2$  & $|0\rangle_{2} \rightarrow |U\rangle_{2}$                                                      &  $\Omega_R/2$ &  $\pi$  \\
2)~Resonate         & $|1\rangle_{a}|g\rangle_{q}|0\rangle_{b} \rightarrow |0\rangle_{a}|g\rangle_{q}|1\rangle_{b}$  &  $g$          &  $\sqrt{2}\pi$  \\
3)~Resonate         & $|1\rangle_{b}|U\rangle_{2} \leftrightarrow |0\rangle_{b}|0\rangle_{2}$                        &  $g_{2}$      &  $2\pi$  \\
4)~Resonate         & $|0\rangle_{a}|g\rangle_{q}|1\rangle_{b} \rightarrow |1\rangle_{a}|g\rangle_{q}|0\rangle_{b}$  &  $g$          &  $\sqrt{2}\pi$  \\
5)~Resonate         & $|U\rangle_{1}|1\rangle_{a} \rightarrow |1\rangle_{1}|0\rangle_{a}$                            &  $g_{1}$      &  $3\pi$ \\
~~~\,Rotate NVE$_2$  & $|U\rangle_{2} \rightarrow |0\rangle_{2}$                                                      &  $\Omega_R/2$ &  $\pi$  \\
\hline
\hline
\end{tabular}\label{Table1}
\end{table}

In step~(1), we tune the transition $|0\rangle_{1}\leftrightarrow
|U\rangle_{1}$ of NVE$_1$ to be resonant with TLR$_a$, which is
similar to the second step  in our state-transfer protocol. The
effective Hamiltonian of the subsystem consisting of NVE$_{1}$ and
TLR$_{a}$ is $H_{s1}=H^{\phi}_{2}$. The subsystem evolves from
$|0\rangle_{1}|0\rangle_{a}$ to  the state
$-i|U\rangle_{1}|1\rangle_{a}$ at the appropriate time
$t=\pi/2g_{1}$, with other states unchanged through the evolution
time.

Meanwhile, we apply a drive field described by
$H_{D}=\Omega_{R}\exp(-i\omega_{2,0}t)S^{+}_{2,0}/2+\textrm{h.c.}$
with the Rabi frequency $\Omega_{R}$ and the frequency
to be the transition frequency of $|0\rangle_{2}
\leftrightarrow |U\rangle_{2}$. The Hamiltonian of the subsystem
composed of NVE$_2$ and TLR$_b$ is
$H_{s2}^{\prime}=H_{b}+H_{2}+H_{b2}+H_{D}$. The Hamiltonian can be
approximately reduced to be
$H_{s2}\approx\frac{\Omega_{R}}{2}(S_{2,0}^{+}+S_{2,0}^{-})$. With
the duration $t=\pi/\Omega_{R}$, the drive field applied on  NVE$_2$
makes it evolve from  $|0\rangle_{2}$ to $-i|U\rangle_{2}$, while
the states of TLR$_b$ and SPQ remain unaltered.

After step~(1), the state of the total system becomes
\begin{eqnarray}    
|\psi\rangle_{1} \!\!\! &=& \!\!\! (-\alpha |U\rangle_{1}|U\rangle_{2}|1\rangle_{a}
-i\beta |U\rangle_{1}|1\rangle_{2}|1\rangle_{a}-i\gamma |1\rangle_{1}|U\rangle_{2}|0\rangle_{a}    \nonumber\\
               \!\!\!  && \!\!\! +\delta |1\rangle_{1}|1\rangle_{2}|0\rangle_{a})\otimes |g\rangle_{q}|0\rangle_{b}.
\end{eqnarray}

In step~(2), we use the same method as that in the third step  in
our state-transfer protocol to achieve the state transfer from
TLR$_a$ to  TLR$_b$. With the effective Hamiltonian
$H_{e2}=H^{\phi}_{3}$ operating for the duration $t=\pi/\sqrt{2}g$,
the system evolves from the state $|\psi\rangle_{1}$ to
\begin{eqnarray}    
|\psi\rangle_{2} \!\!\! &=& \!\!\! (\alpha |U\rangle_{1}|U\rangle_{2}|1\rangle_{b}
+i\beta |U\rangle_{1}|1\rangle_{2}|1\rangle_{b}-i\gamma |1\rangle_{1}|U\rangle_{2}|0\rangle_{b} \nonumber\\
                 && +\delta |1\rangle_{1}|1\rangle_{2}|0\rangle_{b})\otimes|g\rangle_{q}|0\rangle_{a}.
\end{eqnarray}

In  step~(3), we exploit the  Hamiltonian $H_{e3}=H_4^{\phi}$ to
achieve the resonant interaction between TLR$_b$ and the transition
$\vert U\rangle_2\leftrightarrow \vert 0\rangle_2$, similar to the
fourth step in our state-transfer protocol. With the interval
$t=\pi/g_{2}$, the state of the system becomes
\begin{eqnarray}    
|\psi\rangle_{3} \!\!\! &=& \!\!\! (-\alpha |U\rangle_{1}|U\rangle_{2}|1\rangle_{b}
+i\beta |U\rangle_{1}|1\rangle_{2}|1\rangle_{b}-i\gamma |1\rangle_{1}|U\rangle_{2}|0\rangle_{b}  \nonumber\\
               \!\!\! && \!\!\! +\delta |1\rangle_{1}|1\rangle_{2}|0\rangle_{b})\otimes|g\rangle_{q}|0\rangle_{a}.
\end{eqnarray}

Step~(4) is the same as step~(2). By virtue of simultaneous resonant
interactions between the SPQ and the two TLRs, we can obtain the
state
\begin{eqnarray}    
|\psi\rangle_{4} \!\!\! &=& \!\!\! (\alpha |U\rangle_{1}|U\rangle_{2}|1\rangle_{a}
-i\beta |U\rangle_{1}|1\rangle_{2}|1\rangle_{a}-i\gamma |1\rangle_{1}|U\rangle_{2}|0\rangle_{a} \nonumber\\
                \!\!\! && \!\!\! +\delta |1\rangle_{1}|1\rangle_{2}|0\rangle_{a})\otimes|g\rangle_{q}|0\rangle_{b}.
\end{eqnarray}

The last step~(5) is the same as step~(1). A drive pulse with the
duration $t=\pi/\Omega_{R}$ is applied to induce the transition
between $|0\rangle_{2}$ and $|U\rangle_{2}$ of NVE$_2$. Meanwhile, a
resonant interaction between NVE$_1$ and TLR$_a$ lasts for
$g_{1}t=3\pi/2$. Thus an overall c-phase gate between NVE$_1$ and
NVE$_2$ is implemented, leaving the TLR$_a$, TLR$_b$ and the SPQ in
the vacuum and ground states, that is,
\begin{eqnarray}    
|\psi\rangle_{F} \!&=&\!(\alpha |0\rangle_{1}|0\rangle_{2}+\beta
|0\rangle_{1}|1\rangle_{2}
-\gamma |1\rangle_{1}|0\rangle_{2}+\delta |1\rangle_{1}|1\rangle_{2})\;\;\;\;\nonumber\\
                 && \otimes|g\rangle_{q}|0\rangle_{a}|0\rangle_{b}.
\end{eqnarray}

Our simulation on the dynamics of the system with the Hamiltonian in
Eq.~(\ref{state12}) shows that the average fidelity of our c-phase
gate is $98.23\%$ within the operation time $93.87$ ns. Here the
average fidelity is defined as
\begin{eqnarray}              
F_{\textrm{c-phase}}=(\frac{1}{2\pi})^2\int_{0}^{2\pi}\int_{0}^{2\pi}\, _{F}\langle\psi|\rho_{f}^{\textrm{c-phase}}|\psi\rangle_{F}
d\theta_1d\theta_2,\nonumber\\ \label{fidelity2}
\end{eqnarray}
similar to that in Eq.~(\ref{fidelity}). In our simulation, the
parameters are chosen as
$\omega_{a}/2\pi=\omega_{eg}/2\pi=\omega_{b}/2\pi=1.4$ GHz,
$\omega_{1,0}/2\pi=\omega_{2,0}/2\pi=2.08$ GHz for the
large-detuning case, $\omega_{1,1}/2\pi=\omega_{2,1}/2\pi=2.88$ GHz,
$g_{1}/2\pi=16$ MHz, $g_{2}/2\pi=20$ MHz, and $\Omega_{R}/2\pi=50$
MHz. $g/2\pi=104$ (0.5) MHz when we turn  up (down) the coupling
between the SPQ and TLRs.  As an example for the fidelity of our
gate with $\theta_1=\theta_2=\pi/4$, the density operators of the
initial state and the final state are shown in Fig.~\ref{figure3}.
Here, the density matrix is spanned
in the basis  $\{|0\rangle_{1}|0\rangle_{2}|g\rangle_{q},
~|0\rangle_{1}|0\rangle_{2}|e\rangle_{q},
~|0\rangle_{1}|1\rangle_{2}|g\rangle_{q},
~|0\rangle_{1}|1\rangle_{2}|e\rangle_{q},$
$~|1\rangle_{1}|0\rangle_{2}|g\rangle_{q},
~|1\rangle_{1}|0\rangle_{2}|e\rangle_{q},
~|1\rangle_{1}|1\rangle_{2}|g\rangle_{q},
~|1\rangle_{1}|1\rangle_{2}|e\rangle_{q}\}$.

\begin{table}
\caption{Protocol for realization of CNOT gate between NVE$_1$ and NVE$_2$.}
\begin{tabular}{lccc}
\hline
\hline
   Step         & Transition & Coupling & ~Pulse \\ \hline
1)~Resonate     & $|0\rangle_{1}|0\rangle_{a} \rightarrow |U\rangle_{1}|1\rangle_{a}$                           &  $g_{1}$      &  $\pi$    \\
~~~\,Rotate     & $|0\rangle_{2} \rightarrow |U\rangle_{2}$                                                     &  $\Omega_R/2$ &  $\pi$    \\
2)~Resonate     & $|1\rangle_{a}|g\rangle_{q}|0\rangle_{b} \rightarrow |0\rangle_{a}|g\rangle_{q}|1\rangle_{b}$ &  $g$          &  $\sqrt{2}\pi$    \\
3)~Resonate     & $|1\rangle_{b}|U\rangle_{2} \rightarrow |0\rangle_{b}|0\rangle_{2}$                           &  $g_{2}$      &  $\pi$    \\
4)~Resonate     & $|1\rangle_{2} \leftrightarrow |U\rangle_{2}$                                                 &  $\Omega_R/2$ &  $\pi$    \\
5)~Resonate     & $|1\rangle_{b}|U\rangle_{2} \rightarrow |0\rangle_{b}|0\rangle_{2}$                           &  $g_{2}$      &  $\pi$    \\
6)~Resonate     & $|1\rangle_{2} \leftrightarrow |U\rangle_{2}$                                                 &  $\Omega_R/2$ &  $3\pi$   \\
7)~Resonate     & $|0\rangle_{b}|0\rangle_{2} \rightarrow |1\rangle_{b}|U\rangle_{2}$                           &  $g_{2}$      &  $\pi$    \\
8)~Resonate     & $|0\rangle_{a}|g\rangle_{q}|b\rangle_{b} \rightarrow |1\rangle_{a}|g\rangle_{q}|0\rangle_{b}$ &  $g$          &  $\sqrt{2}\pi$    \\
9)~Resonate     & $|U\rangle_{1}|1\rangle_{a} \rightarrow |0\rangle_{1}|0\rangle_{a}$                           &  $g_{1}$      &  $\pi$    \\
~~~\,Rotate     & $|U\rangle_{2} \rightarrow |0\rangle_{2}$                                                     &  $\Omega_R/2$ &  $3\pi$   \\
\hline
\hline
\end{tabular}\label{Table2}
\end{table}

Local resonant interaction and single-qubit operations can also be used to construct the fast CNOT gate on NVEs in the hybrid device. The matrix of the CNOT gate reads
\[U_\textrm{CNOT} = \begin{pmatrix}
0 & 1 & 0 & 0 \\
1 & 0 & 0 & 0 \\
0 & 0 & 1 & 0 \\
0 & 0 & 0 & 1
\end{pmatrix}\]
in the computational two-NVEs basis, that is, $\{|0\rangle_{1}|0\rangle_{2},|0\rangle_{1}|1\rangle_{1},|1\rangle_{1}|0\rangle_{2},|1\rangle_{1}|1\rangle_{2}\}$. The nine steps for the construction of the CNOT gate on two NVEs are shown in Table~\ref{Table2}. It can be implemented with the processes similar to those for our c-phase gate.

\section{GENERATION of CLUSTER STATE IN ONE-DIMENSIONAL and TWO-DIMENSIONAL CIRCUITS}
\label{sec:ClusterState}

\begin{figure}[!h]
\begin{center}
\includegraphics[width=8.5 cm,angle=0]{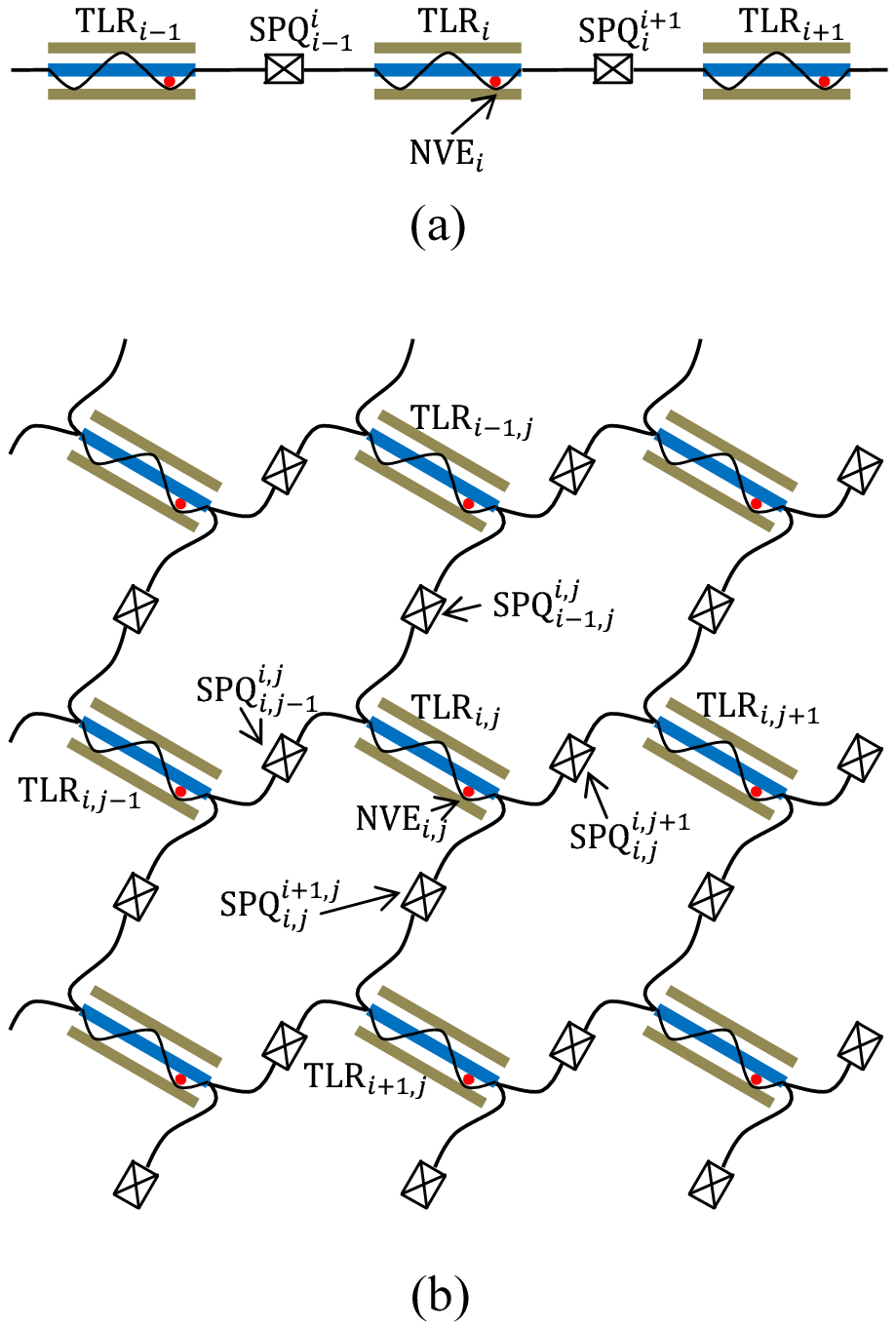}
\caption{(Color online) The schematic layout of generating the large
cluster state based on the 1D circuit chain (a) and the 2D square
grid circuit (b). And the red dot embedded in the TLR represents an
NVE.}\label{figure4}
\end{center}
\end{figure}


Using the c-phase gate, one can construct a two-dimensional (2D)
cluster state, which can be used to realize a one-way quantum
computing \cite{Nielsen1,Wu,Haack,Raussendorf,Nielsen2}. Before
generating a 2D cluster state in a hybrid circuit grid, we try to
implement a one-dimensional (1D) cluster state \cite{Nielsen2} in a
hybrid circuit chain. Now, we will demonstrate in detail how to make
use of the initial state $\prod^{\otimes
n}_{i=1}(|0\rangle_i+|1\rangle_i)/\sqrt{2}$ and our c-phase gate to
generate the large NVE cluster state. In order to realize this
initial state, we can apply two external drive fields to induce the
transitions of $|U\rangle\leftrightarrow|0\rangle$ and
$|U\rangle\leftrightarrow|1\rangle$. In the first step, as shown in
Fig.~\ref{figure4}(a), we divide the NVEs into many pairs
NVE$_{2i-1}$---NVE$_{2i}$ ($i=1,2,\cdots$) and tune the transition
frequencies of $\textrm{SPQ}_{2i}^{2i+1}$ ($i=1,2,\cdots$) to the
largely-detuned regime to form independent pairs in which each is a
subsystem shown in Fig.~\ref{figure1}(a). Then we operate c-phase
gates between NVE$_{2i-1}$ and NVE$_{2i}$ ($i=1,2,\cdots$). After
this step, the state of the system composed of all the NVEs is
\begin{eqnarray}    
\frac{1}{2^n}\prod^{\otimes n}_{i=1}(|0\rangle_{2i-1}+\sigma_{2i}^{z}|1\rangle_{2i-1})(|0\rangle_{2i}+|1\rangle_{2i}),
\end{eqnarray}
where $\sigma_{2i}^{z}$ is the Pauli-Z operator for NVE$_{2i}$. In
the second step, we perform the c-phase gate on NVE pairs
$(2i,2i+1)$ ($i=1,2,\cdots$) as the same as that in the first step.
Then we can prepare the chain in the cluster state
\begin{eqnarray}    
\frac{1}{\sqrt{2^n}}\prod^{\otimes n}_{i=1}(|0\rangle_{i}+\sigma_{i+1}^{z}|1\rangle_{i}),
\end{eqnarray}
where $\sigma_{n+1}^{z}\equiv1$.

Now, we will demonstrate the four steps to generate a 2D cluster
state in the $n\times n$ square grid \cite{Helmer} as shown in
Fig.~\ref{figure4}(b). First of all, as in the 1D case, two sets of
c-phase gates are sequentially performed to prepare the NVEs in each
row into a 1D cluster state
\begin{eqnarray}    
\frac{1}{\sqrt{2^{n^2}}}\prod^{\otimes n}_{i,j=1}(|0\rangle_{i,j}+\sigma_{i,j+1}^{z}|1\rangle_{i,j}),
\end{eqnarray}
where $\sigma_{i,j+1}^{z}$ is the Pauli-Z operator for
NVE$_{i,j+1}$. Second, the same operations are performed on the
columns as on the rows. Then the 2D cluster state is
\begin{eqnarray}    
\frac{1}{\sqrt{2^{n^2}}}\prod^{\otimes n}_{i,j=1}(|0\rangle_{i,j}+\sigma_{i,j+1}^{z}\sigma_{i+1,j}^{z}|1\rangle_{i,j}),
\end{eqnarray}
where $\sigma_{i,n+1}^{z}\equiv\sigma_{n+1,j}^{z}\equiv1$. In fact,
this method can be extended to the general case, i.e.,  to prepare a
$d$D cluster in which $2d$ steps are needed since in each dimension
$2$ steps are required.

\section{DISCUSSION AND SUMMARY}
\label{sec:Discussion}

Recently, the  hybrid quantum system made up of NVEs and
superconducting circuits has been studied for quantum computation
\cite{Kubo2,Yang1,Yang2}. In the system, the coupling strength
between an NVE and a TLR can be enhanced to about $10\sim65$  MHz
\cite{Kubo1,Sandner}, and the NVE can act as either a qubit or  a
good memory because the coherence time of an NV-center is much
longer than that of an SPQ \cite{Jelezko}.

In previous works about hybrid systems, the proposals for the
entanglement or information transfer between two NVEs with the
states $|m_s=0\rangle$ and $|m_s=\pm 1\rangle$ \cite{Yang1,Yang2}
has been studied. To avoid the indirect interaction between the two
NVEs, which can be induced by coupling with the same field mode, we
place these two NVEs in two different TLRs. Moreover, because the
two TLRs are connected by an SPQ with tunable couplings
\cite{Allman,Srinivasan}, the induced interaction between the two
NVEs can be effectively turned on and off. On the other hand, using
the states $|m_s=0\rangle$ and $|m_s=1\rangle$ alone with the fixed
level spacing  leads to the difficulty in operation \cite{Helmer}.
In order to overcome this problem, we construct the fast universal
quantum gate by using the computational states $|m_s=-1\rangle_i$
and $|m_s=+1\rangle_i$ in combination with the third auxiliary
energy level $|m_s=0\rangle_i$, which  gives us more freedom to
achieve quantum information processing.

In 2012, in an interesting work by Chen \emph{et al.}
\cite{fengmang}, the operation time of quantum state transfer, from
the initial state
$(\alpha\vert0\rangle_{\textrm{NVE}_{1}}+\beta\vert1\rangle_{\textrm{NVE}_{1}})\vert0\rangle_{\textrm{NVE}_{2}}$
to the final state
$\vert0\rangle_{\textrm{NVE}_{1}}(\alpha\vert0\rangle_{\textrm{NVE}_{2}}-i\beta\vert1\rangle_{\textrm{NVE}_{2}})$,
needs only $30$ ns with coupling strength between NVE and
superconducting circuits about $70$ MHz. We remark that our proposal adopts a different final state with respect to theirs. If we choose the state transfer from
$(\alpha\vert0\rangle_{\textrm{NVE}_{1}}+\beta\vert1\rangle_{\textrm{NVE}_{1}})\vert
U\rangle_{\textrm{NVE}_{2}}$  to the same final state as theirs
with the same coupling between NVE and superconducting circuits,
the whole procedure in our proposal will reduce to four steps and the
whole operation time is significantly reduced to $13.41$ ns with the fidelity
about $96.88\%$. In addition, different from the proposal proposed by
Yang \emph{et. al.} \cite{Yang2} in 2012 for the state transfer
between two NVEs within $400$ ns by using the global resonance on the
whole hybrid system, our protocol for this task requires merely $70.60$ ns
by using the local resonance between an NVE (the SPQ) and TLRs.
Another advantage of the local resonance is that by virtue of the local resonance we can
construct a multi-dimensional cluster state with only a few steps.

Resonance operation between an artificial atom and a cavity is one
of the fast quantum operations. The resonance operation between an
NVE and a superconducting resonator can be completed with a very
high fidelity about 97\% \cite{Yang1}. The resonance operation
between an SPQ and a superconducting resonator can also be achieved
with a very high fidelity, as shown in
Refs.~\cite{Haack,You2,Sillanp}. The coupling strength between the
qubit and the resonator can be achieved as high as 100 MHz
\cite{Allman}, which suggests the quantum information transfer from
resonator $a$ to resonator $b$ can be achieved within a very short
time, compared to the decoherence time. The main factor which limits
the operation time of our c-phase gate is the interaction between
the NVE and the resonator. Since the couplings between NV-centers
and resonator are quasi-homogeneous, the coupling strength between
the collective mode and the resonator has been enhanced by
$\sqrt{N}$ \cite{Song,Ai2}. In our simulations, we do not consider
the decoherence and leakage mechanisms of the hybrid system, due to
the short operation time $93.87$ ns as compared to the coherence
times of NV-center $\sim10^{-3}$s \cite{Jelezko,Choi} and the SPQ
$\sim10^{-5}$s \cite{Xiang,Yu}, and the large quality factor of the
superconducting resonator $>10^6$ \cite{Devoret,Chen2,Leek,Megrant}.
We remark that the quantum dynamics given by quantum master equation
\cite{Breuer02} which takes decoherent effects into account and can
be solved by quantum Monte Carlo approach
\cite{Dalibard92,Piilo08,Ai14,Ai13} will be essentially very close
to the present result. With the help of the short operation time of
the c-phase gate, we can effectively construct the one-way quantum
computation. Due to the long coherence time of the NVE, our one-way
quantum computation based on the NVE has a longer life time than the
effective scheme by Wu \emph{et al}. \cite{Wu} based on the 1D
superconducting resonators.

In summary, we have proposed an effective scheme for the state
transfer between two remote NVEs and that for the fast c-phase gate
on them. Our hybrid system  consists of two distant NVEs coupled to
separated high-$Q$ TLRs, which are interconnected by an SPQ. The
quantum state transfer and the c-phase gate are implemented by using
local resonant interaction between the NVE and the resonator, and
the single-qubit operation on the NVE, not global resonance
\cite{Yang1,Yang2}. The fidelity of our quantum state transfer is
$99.65\%$ within a short operation $70.60$ ns. The fidelity of our
c-phase gate is $98.23\%$ within a short operation time of $93.87$
ns. Assisted by our c-phase gate, we propose a scheme to generate a
two-dimensional cluster state on distinct NVEs in a square grid
based on the above hybrid quantum system interconnected by the
charge qubits. In this hybrid system, we can construct a one-way
quantum computation with long coherent time in comparison with that
based on the pure superconducting circuit system.

\section*{ACKNOWLEDGMENTS}

This work was supported by the National Natural Science Foundation
of China under Grant Nos.11174039 and 11474026,  NECT-11-0031, and
the Youth Scholars Program of Beijing Normal University under Grant
No. 2014NT28.

\end{document}